# RCMNet：A deep learning model assists CAR-T therapy for leukemia

Ruitao Zhang[1,2], Xueying Han[3], Ijaz Gul[1,2], Shiyao Zhai[1,2], Ying Liu[4], Yongbing Zhang[5], Yuhan Dong[1,2], Lan Ma[1,2], Dongmei Yu[6], Jin Zhou[3], Peiwu Qin[1,2]

[1]Institute of Biomedical and Health Engineering, Tsinghua Shenzhen International Graduate School

[2]Precision Medicine and Public Health, Tsinghua-berkeley Shenzhen Institute

[3]The First Affiliated Hospital of Harbin Medical University

[4]Food Inspection & Quarantine Center, Shenzhen Custom

[5] Department of Computer Science, Harbin Insitute of Technology, Shenzhen

[6]School of Mc chanical, Electrical & Information Engineering, Shandong University

## Abstract

Acute leukemia is a type of blood cancer with a high mortality rate. Current therapeutic methods include bone marrow transplantation, supportive therapy, and chemotherapy. Although a satisfactory remission of the disease can be achieved, the risk of recurrence is still high. Therefore, novel treatments are demanding. Chimeric antigen receptor-T (CAR-T) therapy has emerged as a promising approach to treat and cure acute leukemia. To harness the therapeutic potential of CAR-T cell therapy for blood diseases, reliable cell morphological identification is crucial. Nevertheless, the identification of CAR-T cells is a big challenge posed by their phenotypic similarity with other blood cells. To address this substantial clinical challenge, herein we first construct a CAR-T dataset with 500 original microscopy images after staining. Following that, we create a novel integrated model called RCMNet (ResNet18 with CBAM and MHSA) that combines the convolutional neural network (CNN) and Transformer. The model shows 99.63% top-1 accuracy on the public dataset. Compared with previous reports, our model obtains satisfactory results for image classification. Although testing on the CAR-T cells dataset, a decent performance is observed, which is attributed to the limited size of the dataset. Transfer learning is adapted for RCMNet and a maximum of 83.36% accuracy has been achieved, which is higher than other SOTA models. The study evaluates the effectiveness of RCMNet on a big public dataset and translates it to a clinical dataset for diagnostic applications.

## 1. Introduction

Leukemia is a common hematopoietic malignant disease, which is hard to cure due to its malignant proliferation in the human body. According to global cancer statistics, 311,594 deaths and 474,519 new cases of leukemia have been reported worldwide in 2020 (*1*), and the number is rising gradually per year. The number of males reaches up to 269,503, which is higher than 205,016 for females and the ratio can be up to 1.31 (*1*). Leukemia is categorized into two types, acute leukemia malignantly proliferating from the primitive white blood cells such as hematopoietic stem cells, and chronic leukemia malignantly proliferating from mature white blood cells. The worsening progress is faster and the death rate is higher for acute leukemia (*2*), leading to substantial therapeutic challenges. Generally, the methods to treat acute leukemia are bone marrow transplantation, supportive therapy, and chemotherapy. Although bone marrow transplantation is an acceptable approach to treat leukemia, some reports have demonstrated that once bone marrow transplantation is not enough, more transplants are required (*3, 4*). The research shows that bone marrow transplantation occurs with a high relapse incidence and low overall survival (*5*). In addition, cure rates are related to the age stages, and the cure rate of childhood acute lymphoblastic leukemia can be up to over 90% (*6*). Patients of some subtypes have lower survival rates as immune cells cannot target the cancer cells. Novel therapeutic modalities to treat leukemia patients are urgently needed.

CAR-T cell immunotherapy has shown the potential to address this challenge. A CAR-T cell is a type of artificial T cell, which is obtained by extracting the T cells from the leukemia patient and then modifying them by inserting a sequence of genes that can express receptor-recognizing tumor antigen and co-stimulatory molecules. The CAR-T cells are then injected back into the patient. Figure 1 illustrates the procedures of therapy and detection. For patients with leukemia, CAR-T cells can target the tumor cells specifically and consistently by proliferation to kill the abnormal cells compared with the drugs from chemotherapy that attacks cells indistinguishably. Hence, CAR-T immunotherapy is deemed to be one of the most promising ways to cure leukemia and even other cancers. Clinical trials have shown that CAR-T can achieve a remission rate of 70–90% (*7, 8*). The complete elimination of blood cancer cells by CAR-T immunotherapy has been demonstrated, though the long-term evaluation is unknown (*9, 10*).

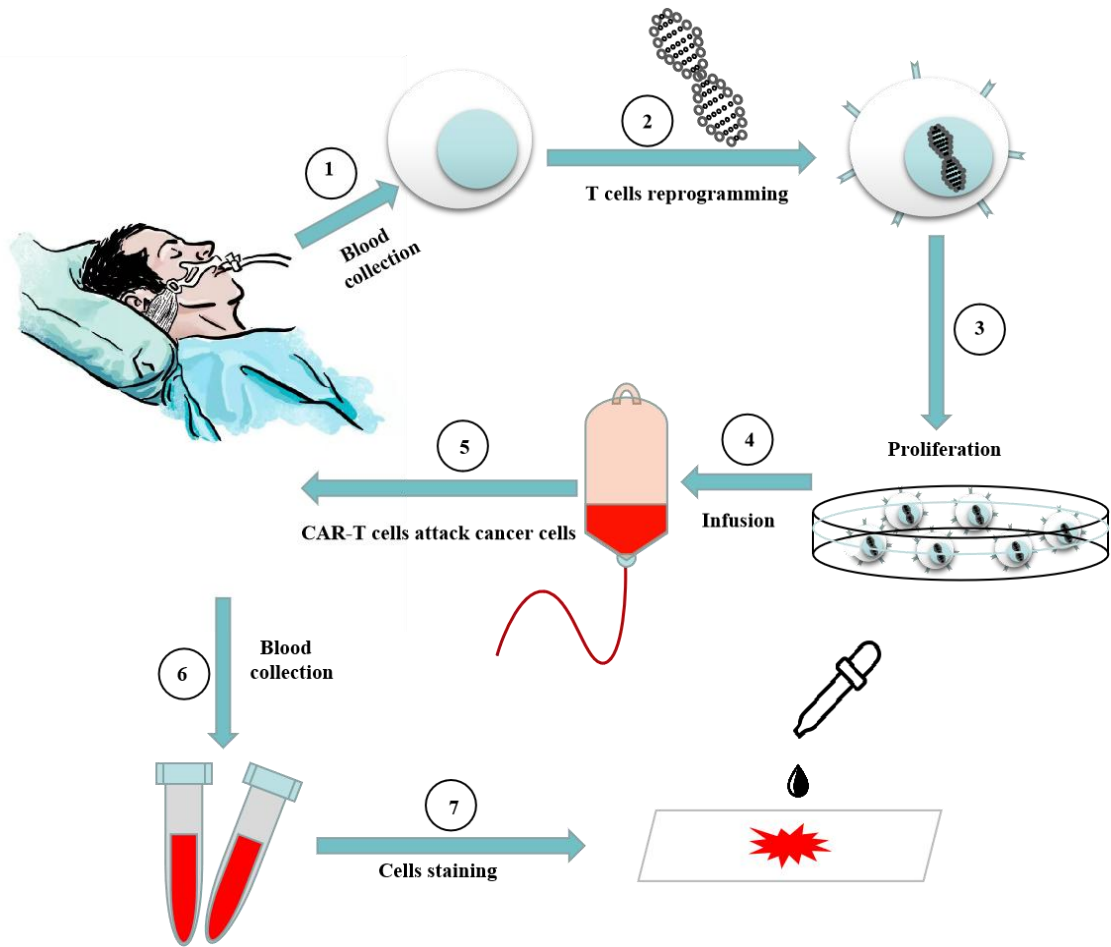

Figure 1: The flow chart of CAR-T cell therapy and CART staining for characterization.

Morphological identification of CAR-T cells is very important for blood diseases. Because of the individual differences, the patients receiving CAR-T cell therapy may occurring the side-reactions, which it can be severely debilitating or fatal. It is extremely significant to instruct the clinical treatment by monitoring the development of immune cells in detail and real time. A hematology analyzer can pinpoint the abnormal phenotypic cells rapidly. However, the automated analysis does not show better performance compared with the manual identification, which is indispensable for the successful diagnosis of leukemia (*11*). Although CAR-T immunotherapy is a promising treatment to cure leukemia, limited blood morphologists can identify the morphological features of the CAR-T cells precisely because of their heterogeneous nature in different patients and the similar phenotypes with other cells. Figure 2 shows the examples of CART cells and interfering blood cells with similar morphology, which is difficult to be distinguished by human eyes. Furthermore, the biochemical analysis or training a novice blood morphologist who can recognize the CAR-T cell is time-consuming and expensive. The phenotype identification of CAR-T cells is crucial for disease prognosis. A survey has reported the partial characteristics of CAR-T cells, such as the large multinucleated forms (*12*). However, there are no further reports on CAR-T cell characterization.

In recent years, deep learning (DL) has enticed substantial research attention by constructing models to assist doctors in diagnosing a variety of diseases. It has shown great success in medical cell image classification (*13-15*). For image analysis, a reliable dataset is of paramount importance; the morphological observation is crucial for blood disease diagnosis and recognition of CAR-T cells is significant for the prognosis of patients after receiving CAR-T therapy. Introducing a system for morphological studies of CAR-T cells can be a good addition to the leukemia treatment modalities. This study comprises of two parts: constructing a CAR-T cell dataset and proposing a novel DL model to distinguish CAR-T cells precisely. The contributions of our study are as follows:
- Construct the first CAR-T cell dataset from patients with leukemia with ethical approval.
- Design and implement a new multi-attention network for the CAR-T cell classification, which combines CNN with self-attention and apply it to CAR-T cell classification for the first time.

The organization of the paper is as following: Section 2 shows relative work on cell classification. Section 3 illustrates the data collection and method of building our model. Section 4 presents the result and discussion. Lastly in section 5, the conclusion and future work are presented.

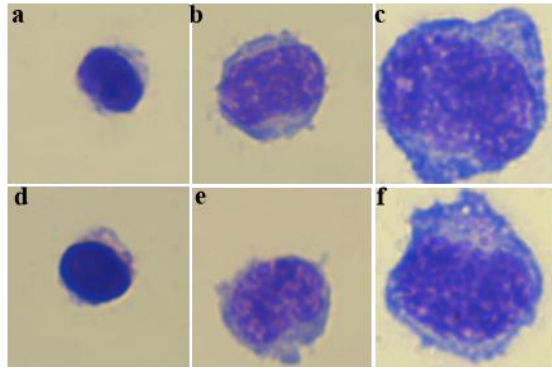

Figure 2: Comparison between images of CAR-T cells (a-c) and other cells (d-f) with various shapes and sizes.

## 2. Relative work

## 2.1 CNN

Computer vision (CV) has shown increased applications in autonomous technology and advanced surveillance systems (*16*). In recent years, medical image processing has enticed substantial research attention due to increasing shared medical resources, which can assist physicians in diseases diagnosis and capture potential features that physicians rarely notice (*13*). In contrast to doctors, CV can work continuously and efficiently, maintaining decent accuracy. The most classical method is machine learning (ML). Before ML classification, pre-processing procedure extracts features with key information, such as principal component analysis (PCA) (*17*), independent component analysis (ICA) (*18*), and linear discriminant analysis (LDA) (*19*). However, it is unable to capture all prominent features and thus loses some crucial information, resulting in lower accuracy. The classification methods can be divided into two categories: supervised learning and unsupervised learning. Supervised learning techniques include support vector machine (SVM) (*20*), k-nearest neighbor (KNN) (*21*), classification and regression tree (CART) (*22*), etc. Among the unsupervised learning approaches, the K-Means (*23*) is the most common approach. ML cannot obtain higher accuracy without sufficient information. Therefore, alternative methods are desired to overcome this challenge.

With the development of computer technology, novel hardware and software can provide and support larger memory and higher hashrate for DL training. The appearance of DL resolves the limitations of ML. CNN is the major player in DL, which consists of many convolutional layers that contain convolutional kernels of various sizes, in an end-to-end learning schema. Via convolutional kernels, CNN can extract a mass of features randomly and assemble the key features for downstream task. The application of CNN in medical image processing is appealing to both researchers and doctors. For cell classification, outer phenotype and inner structure are two frequently-used input features. The reports about cell recognition include blood cells, cancer cells, and other cell types (*22, 24-26*).

Many models have been adapted and created for cell recognition. A hybrid model that combines transfer learning with generative adversarial networks (GANs) increases the accuracy of a small dataset with staining-free cancer cells. They collect the optical path delay maps from low-coherence off-axis holography as input and pretrain the GANs with sperm cells before training their dataset. The accuracy of the model is 99%, which outperforms the single GANs or MobileNet with transfer learning (*24*). Another report achieves an accuracy of 99.54% by combining DL and support vector machine (SVM) to classify the sickle cells and normal blood cells with transfer learning and data enhancement (*25*). A deeply supervised residual network can classify human epithelial-2 cells with the accuracy of 99.98% (*26*). For the cellular inner structure, a study shows that DL can classify the *Cercopithecus aethiops* monkey kidney cells based on the microtubule networks, and it shows better results compared with the human expert (*27*). Another research reveals that CNN can distinguish the normal and cancer cell in the breast by recognizing discrepancies in the actin cytoskeleton structures that can be served as a supererogatory diagnostic marker, whose performance is outperforming the human expert (*13*).

## 2.2 CNN and Transformer

CNN owns powerful inductive biases, such as local correlation and weight sharing, which improve the accuracy and effectiveness. However, it limits the performance upper bound of the model as well. Although deeper CNN can weaken the effects of limited receptive field and long-range dependence, more complex and larger CNN is needed, which increases the difficulty of training. The Transformer performs well in global correlation, which is extensively used in natural language processing (NLP) and has achieved great success. However, Transformer training is time-consuming and requires an extremely large dataset and high computational memory. Significant efforts have been made to combine CNN and the Transformer to balance each other to get better results. Vit (*28*) crops several patches from image and reshapes the patches following the word embedding as input. The author maintains the Transformer structure and changes the input that imitates the word embedding. Bello *et al*. propose the hybrid model with CNN and Transformer, named AA-ResNet with 77.7% accuracy on ImageNet classification (*29*). Nevertheless, a comprehensive Transformer is complex and less flexible to be transferred to image processing from text processing. Bottleneck Transformer (Bot) block (*30*) utilizes the core self-attention from Transformer to replace the middle convolutional layer of the last blocks from ResNet50. Although the model structure doesn't change too much, it achieves higher accuracy compared with the traditional CNN.

## 2.3 CAR-T dataset

To the best of our knowledge, no CAR-T cell dataset is available so far. The establishment of the CAR-T database requires the collaboration of experienced morphologists with special expertise. In this work, the CAR-T cell dataset, for the first time, is constructed that can be used as the baseline and reference for the later CAR-T cell dataset construction and research.

## 3. Methodology

## 3.1 dataset

We train and test our model and a few popular models for cell recognition on two datasets with different cell types. The first one is a common dataset for microscopic peripheral blood cells and the other one is our dataset for CAR-T cells obtained with ethical approval and patient notification letter.

### 3.1.1 Peripheral blood cells (PBC) dataset

The PBC dataset is published by Acevedo *et al.* in 2020 (*31*), which is one of the largest and most complete available datasets about blood cells. The dataset is collected by the core laboratory at the hospital clinic of Barcelona with the analyzer Cella Vision DM96, where the May Grünwald-Giemsa stain (*32*) is used to stain the cells in the autostainer Sysmex SP1000i. The PBC dataset contains eight cell types, including neutrophils, eosinophils, basophils, lymphocytes, monocytes, immature granulocytes (IGs), erythroblasts, and platelets (Figure 3). There are three subtypes of IG consisting of promyelocytes, myelocytes, and metamyelocytes, which increase the difficulty of classifying the immature granulocytes with other cells. The total number of images is 17,092. The details of the dataset are showed in Table 1. All of those cells are labeled by professional clinical pathologists and the image size is 360×363 pixels. Each image contains one cell. For the PBC dataset, to minimize the cell type imbalance that may influence the learning effectiveness of the model, the number of lymphocytes is regarded as standard and other cells type are truncated to the same number and all selected images are chosen randomly from the corresponding categories. Table 1 presents the descriptive details of all types of cells.

Table 1: The cell type, number, percentage, and the number of images used for the train and test set for the PBC dataset.

| Cell type | Number | Percentage | Train Set | Test Set |
|---|---|---|---|---|
| Neutrophils | 3329 | 19.48% | 971 | 243 |
| Eosinophils | 3117 | 18.24% | 971 | 243 |
| Immature granulocytes (Metamyelocytes, Myelocytes and Promyelocytes) | 2895 | 16.94% | 971 | 243 |
| Platelets (Thrombocytes) | 2348 | 13.74% | 971 | 243 |
| Erythroblasts | 1551 | 9.07% | 971 | 243 |
| Monocytes | 1420 | 8.31% | 971 | 243 |
| Basophils | 1218 | 7.13% | 971 | 243 |
| Lymphocytes | 1214 | 7.10% | 971 | 243 |
| Total | 17,092 | 100% | 7768 | 1458 |

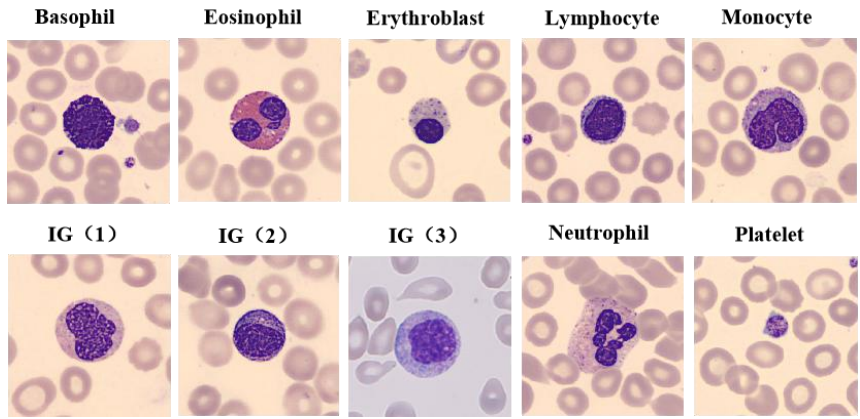

Figure 3: Representative images from 8 classes of PBC dataset showing Basophils (a), Eosinophils (b), Erythroblasts (c), Lymphocytes (d), Monocytes (e), IGs (f-h), Neutrophils (i), Platelets (j). IG1, IG2 and IG3 represent the metamyelocytes, myelocytes, and promyelocytes respectively.

### 3.1.2 CAR-T cell dataset

Six patients with acute lymphoblastic leukemia have received CAR-T therapy and clinical biopsy from the first affiliated hospital of Harbin medical university. The blood samples are collected after the CAR-T therapy within several days or weeks. All patients have signed an informed agreement letter to provide blood samples for this study. The protocol gains ethical approval by the hospital ethical committee. All blood samples, which are gathered from the patients receiving CAR-T cell therapy, are stained by May Grünwald-Giemsa (*32*) and the CAR-T cells have been confirmed by immunostaining for CAR-T specific markers (*12*). To collect CAR-T images, we use wide-field microscopy with a 100x oil immersion lens (Leica, DM500). The size of each image is 384×384 pixels with only one cell. Because of the complexity and scarcity of blood, we label the CAR-T cells with the help of professional blood morphologist. For our dataset, there are two categories of cells, including the CAR-T cells and the other cells. We collect 250 pictures per class. The assignment ratio is 8:2 for each class, which means there are 200 images for the training set and 50 samples for the test set. The dataset description is showed in table 2.

Table 2: CAR-T cell dataset.

| The Type of Cell | Number | Percentage | Train Set | Test Set |
|---|---|---|---|---|
| CAR-T cell | 250 | 50% | 200 | 50 |
| Normal cell | 250 | 50% | 200 | 50 |
| Total | 500 | 100% | 400 | 100 |

### 3.2 Data augmentation

Because a limited number of images may lead to the model overfitting. Dataset expansion is an efficient methodology to decrease overfitting. A similar strategy has been reported earlier (*33*). We utilize data augmentation to increase the size of the limited dataset with rotations and flips including three rotations: 90°, 180°, and 270°, and two flips: vertical and horizontal flip. The final size of the CAR-T dataset expands from 500 to 3000 images with approximately 1500 images per class to balance the dataset (Table 3)

Table 3: CAR-T cell dataset after data augmentation.

| The Type of Cell | Number | Percentage | Train Set | Test Set |
|---|---|---|---|---|
| CAR-T cell | 1500 | 50% | 1200 | 300 |
| Normal cell | 1500 | 50% | 1200 | 300 |
| Total | 3000 | 100% | 2400 | 600 |

### 3.3 Model

The schematic of the proposed model is illustrated in Figure 4. PBC and CAR-T datasets are used as input separately. The model output is the cell recognition based on the features extracted after a series of convolutional operations. ResNet18 serves as the backbone with two inserted blocks consisting of Convolutional Block Attention Module (CBAM) and Multi-Head Self-Attention (MHSA) which are two key attention blocks in our model. The details of the model are described in the following section.

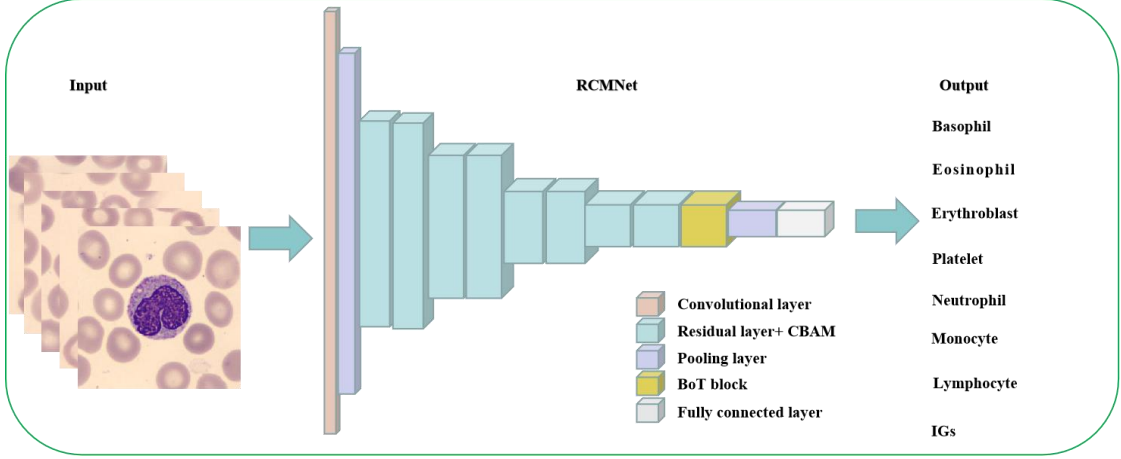

Figure 4: RCMNet schematic. The microscopy images (360×363 or 384×384) are utilized as input. CBAM and BoT block are inserted into ResNet18. The output is the result of cell classification.

### 3.3.1 Residual Neural Network (ResNet)

ResNet is a common and classical neural network in computer vision (*34*). Researchers usually use ResNet as the backbone and modify it for different purposes. There are plenty of ResNet variants, such as ResNeXt (*35*), DenseNet (*36*), and MobileNet (*37*). The unique property of ResNet is the shortcut connection to maintain the original features while addressing the challenge of exploding gradient and vanishing gradient. ResNet is adapted as the backbone of our model, composed of a series of residual blocks. The functions depicting learning process are showed as below:

$$x_{l+1} = Relu(x_l + f(x_l, w_l)) \tag{1}$$

where $x_l$ and $x_{l+1}$ represent the input for the $l_{th}$ layer and output for the $(l+1)_{th}$ layer, $Relu(\cdot)$ is the rectified linear unit function, $f(\cdot)$ represents the residual mapping function, $w_l$ is the parameters for the $l_{th}$ layer.

Based on Equation 1, the cumulative operations (Equation 2) up to the $L_{th}$ layer can be represented as the following:

$$x_L = x_l + \sum_{i=l}^{L-1} f(x_i, w_i). \tag{2}$$

where $x_L$ represents the output for the $L_{th}$ layer, which compress the middle layers by summing the shallow residual block.

### 3.3.2 CBAM

CBAM is an attention mechanism proposed by Woo *et al.* (*38*), derived from Squeeze-and-Excitation Networks (SENet) (*39*). Compared with the common model modification, such as increasing the depth and width, CBAM is a lightweight module, which doesn't consume too much computational memory to achieve higher accuracy. CBAM can be easily inserted into any layer because of the module characteristic. We incorporate CBAM into each residual block of our model. Figure 5 shows the combination of residual block and CBAM module, which is introduced after two convolutional layers. There are two attention modules including the channel attention module and the spatial attention module, where channel attention module is ahead of the spatial attention. Channel attention can be regarded as the feature extractor for each channel. According to fully connection calculating the assigned weight of each channel, the significant features, such as texture, outline, etc, will be integrated and compressed to the size of 1×1×n (n represents the number of classifications) after average pooling and max pooling. The functions for channel attention (Equation 3-5) are showed as below:

$$\varphi_a = F_c\left(Relu\left(F_c(avg(m))\right)\right) \quad (3)$$

$$\varphi_m = F_c\left(Relu\left(F_c(max(m))\right)\right) \quad (4)$$

$$\varphi = \sigma(\varphi_a + \varphi_m) \quad (5)$$

where $m$ is input, $avg(\cdot)$ is average pooling based on width, $max(\cdot)$ is max pooling based on height, $\varphi_a$ is the output after MLP for average pooling feature, $\varphi_m$ is the output after MLP for max pooling, $F_c(\cdot)$ is fully connection, $\sigma(\cdot)$ is the sigmoid function, $\varphi$ is the output of channel attention.

For spatial attention, it calculates the inner relationship between pixels to decide the focus on the image. Using convolution with a 7×7 convolutional kernel, the features after max pooling and average pooling will be compressed to the size of W×H×1 (W represents the width, H is the height) to determine the assignment of spatial weight. The function for special attention (Equation 6) is performed as following:

$$\varepsilon = \sigma\left(conv\left(conc(avg(m), max(m))\right)\right) \quad (6)$$

where $\varepsilon$ represents the output of special attention, $conv(\cdot)$ is the 7×7 convolutional kernel, $conc(\cdot)$ stands for the concatenation between average pooling and max pooling.

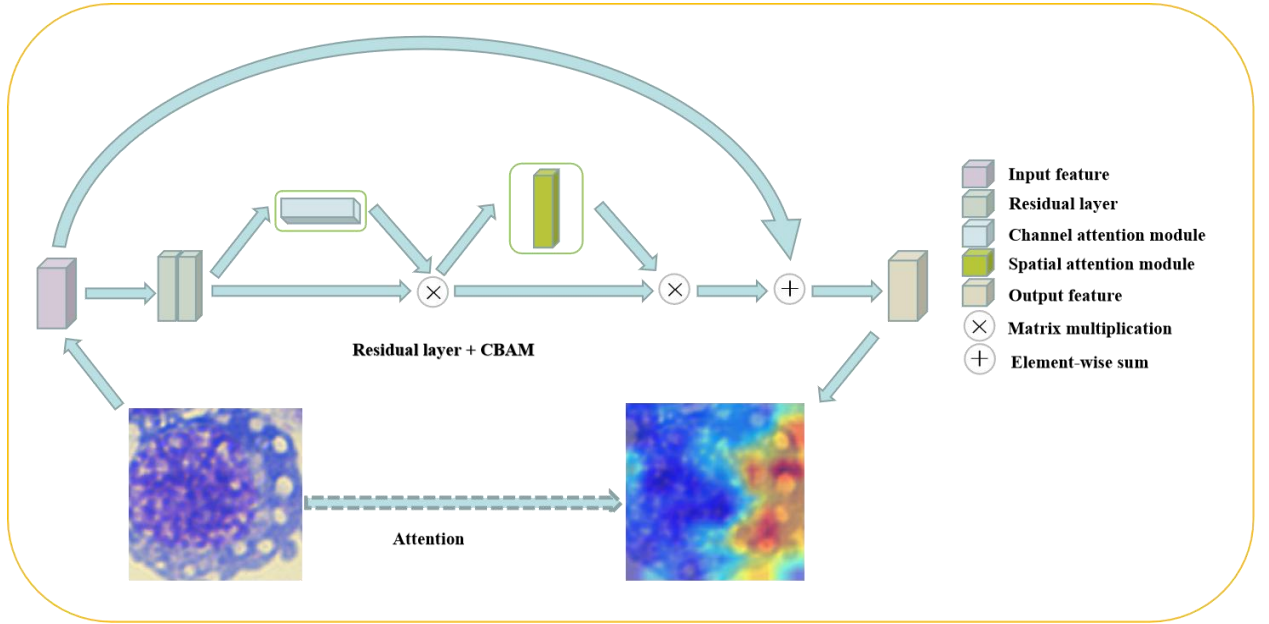

Figure 5: The combination between residual block and CBAM.

### 3.3.3 MHSA

The Bottleneck Transformer Network (BoTNet) is proposed by Srinivas *et al*. (*30*). The middle convolutional layer of the last bottleneck block can be replaced by MHSA in BoT block (*40*), which is commonly used in NLP. The structure of the BoT block is showed in Figure 6a. (*40*). In contrast to MHSA for CNN, Transformer with MHSA utilizes layer normalization, single non-linearity, Adam optimizer, and output projection, while the BoT block uses batch normalization, three non-linearities, SGD optimizer (*30*), and there is no output projection in BoT block. 2D relative position encoding (*41*) is applied to the MHSA of the BoT block to pinpoint the position for each pixel. Figure 6b shows the details of self-attention. Because Transformer can consume a large amount of computational memory, insertion of the BoT block to the last layer of CNN decreases the memory consumption. Inspired by the BoTNet design, we insert the BoT block with 4-heads MHSA after the final layer of the ResNet18. The structure is showed in Figure 6.

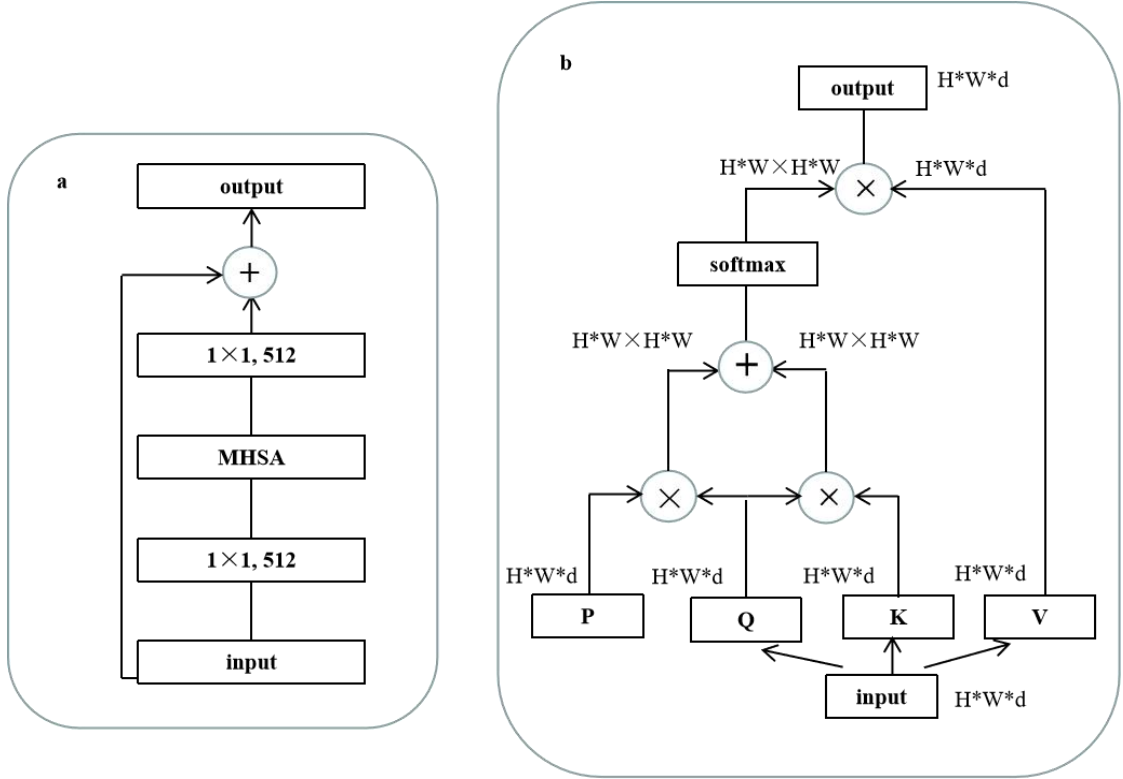

Figure 6: Diagram illustrating the BoT block. The structure of BoT block (a) and self-attention (b). Q, K, V, P represent query, key, value, and position embedding separately. ⊗ represents matrix multiplication and ⊕ represents element-wise sum.

## 3.4 Transfer Learning

Because the doctors have no time to label large amounts of medical images which is time-consuming and there are not enough images in most cases, limited medical data are acquired. Small dataset usually doesn't offer efficient information for the training model leading to underfitting. Transfer learning alleviates the impact of small data size, which is a promising method to increase the model accuracy. Transfer learning transfers the weights learned in previous training with the large public dataset to the target domain, accelerating the learning efficiency without training from scratch. The precondition of transfer learning is similar to data distributions such that prior knowledge can be transferred to the novel model. Raghu *et al*. (*42*) has verified that the transfer learning with ResNet50 shows better result for medical image processing on a more complex model. In this study, transfer learning is adopted because there are limited images for the CAR-T dataset. For RCMNet, the pre-train model is trained on the PBC dataset. Due to the similarity between the PBC dataset and the CAR-T dataset, all weights learned with the PBC dataset except a fully connected layer are transferred and re-trained on the CAR-T dataset. The parameters from the fully connected layer are initialized randomly. Finally, the pre-train model is re-trained on the CAR-T dataset by freezing weights from all layers without the fully connected layer and fine-tuning the weight of the fully connected layer. The schematic is showed in Figure 7.

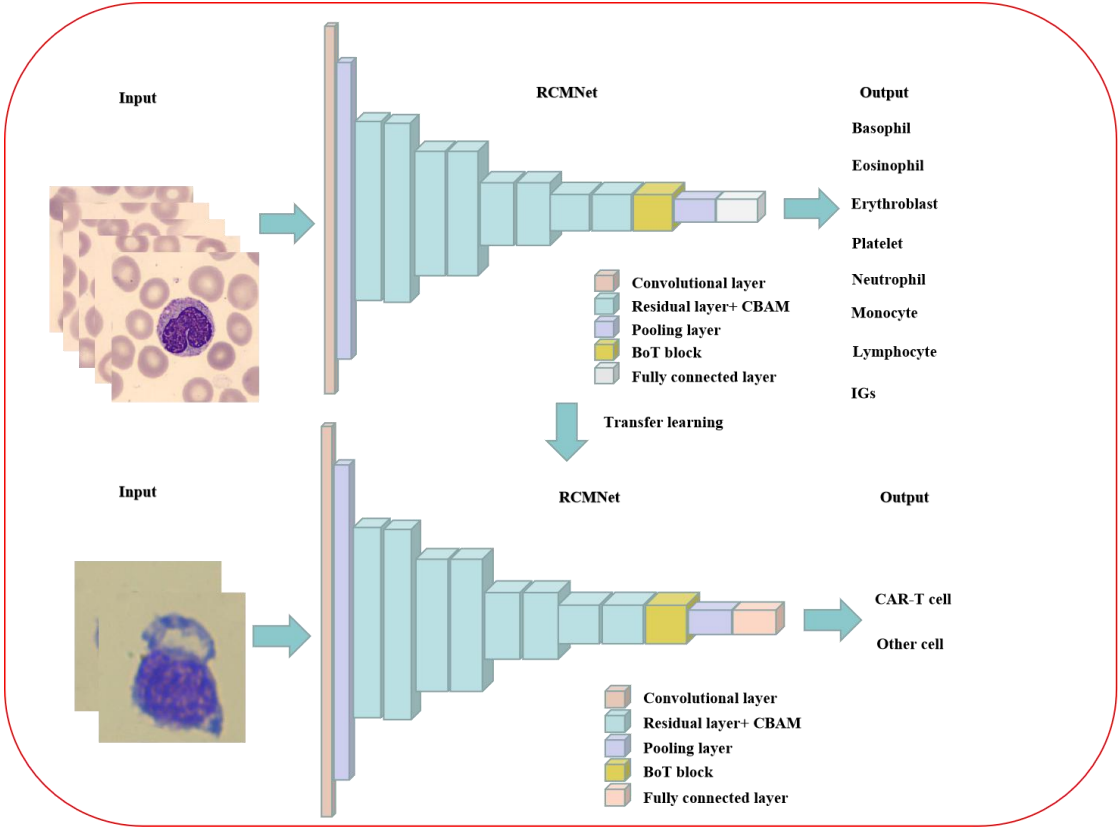

Figure 7: the schematic of transfer learning.

## 3.5 Gradient-weighted Class Activation Mapping (Grad-CAM)

Class activation mapping (CAM) (*43*) visualizes deep learning features, proposed in 2016, which generates the thermodynamic region and illustrates the importance of relevance for decision making in deep learning. Grad-CAM (*44*) doesn't change structure or retrain the model and calculate the gradient of $A_{mn}^i$ according to backpropagation. We obtain the average gradient value by global average pooling for each feature map, which is the $\alpha_i^c$. $\alpha_i^c$ times the corresponding feature map before $Relu$ to get the final Grad-CAM. The functions for Grad-CAM (Equation 7-8) are showed as below:

$$\alpha_i^c = \frac{1}{Z}\sum_{m=1}^{w}\sum_{n=1}^{h}\frac{\partial S_c}{\partial A_{mn}^i} \quad (7)$$

$$L_G^c = Relu\left(\sum_i \alpha_i^c A^i\right) \quad (8)$$

$\alpha_i^c$ represents the sensitivity of the $i_{th}$ channel of feature map in the final convolutional layer; $i$ represents the index of the channel; $c$ represents the index of classification; $A$ represents the feature map from the final convolutional layer; $S$ represents the probability vector; $Z$ represents the size of feature ma; $m\ and\ n$ represent the height and width; $Relu$ represents the activation function; $L_G^c$ represents the Grad-CAM.

## 4. Results

To evaluate the performance of our proposed model, we test the model with two datasets: the PBC dataset and the novel CAR-T dataset. PBC dataset is the benchmark to evaluate the capability of our model with available models and we create an original CAR-T dataset to demonstrate the applicability of our model for the clinical challenges. We compare the performance of our model with other common models (e.g., ResNet, AlexNet, and VGG19). Meanwhile, various ablation experiments are designed to demonstrate the effectiveness of our method on the PBC dataset. Top-1 and Top-5 accuracy levels are reported. All training and testing tasks are operated on the NVIDIA GeForce RTX 2080 SUPER.

## 4.1 Image classification for the PBC dataset

### 4.1.1 Ablation study

To evaluate the effectiveness of the final architecture of the model, we conduct the ablation experiment. ResNet18 is the backbone and all modifications are based on the ResNet18. All the details of the model architecture are explained in Table 4. The designed ResNet18 variant are the model without CBAM and MHSA (ResNet18), the model with CBAM but without MHSA (ResNet18-M), the model with MHSA but without CBAM (ResNet18-C), and the model with MHSA and CBAM (RCMNet).

Table 4: Comparison among various models: ResNet18, ResNet18-M, ResNet18-C and RCMNet. ResNet18 is the backbone. Resnet18-C and ResNet18-M are the variants from ResNet18. RCMNet is our proposed model.

|  | ResNet18 | ResNet18-C | ResNet18-M | RCMNet |
|---|---|---|---|---|
| Conv1 | 7×7 conv,64, Stride2 | | | |
|  | 3×3 Max Pooing, Stride2 | | | |
| Layer1 | [3×3 conv, 64<br>3×3 conv, 64] × 2 | [3×3 conv, 64<br>3×3 conv, 64<br>CBAM, 64] × 2 | [3×3 conv, 64<br>3×3 conv, 64] × 2 | [3×3 conv, 64<br>3×3 conv, 64<br>CBAM, 64] × 2 |
| Layer2 | [3×3 conv, 128<br>3×3 conv, 128] × 2 | [3×3 conv, 128<br>3×3 conv, 128<br>CBAM, 128] × 2 | [3×3 conv, 128<br>3×3 conv, 128] × 2 | [3×3 conv, 128<br>3×3 conv, 128<br>CBAM, 128] × 2 |
| Layer3 | [3×3 conv, 256<br>3×3 conv, 256] × 2 | [3×3 conv, 256<br>3×3 conv, 256<br>CBAM, 256] × 2 | [3×3 conv,256<br>3×3 conv, 256] × 2 | [3×3 conv, 256<br>3×3 conv, 256<br>CBAM, 256] × 2 |
| Layer4 | [3×3 conv, 512<br>3×3 conv, 512] × 2 | [3×3 conv, 512<br>3×3 conv, 512<br>CBAM, 512] × 2 | [3×3 conv,512<br>3×3 conv, 512] × 2 | [3×3 conv, 512<br>3×3 conv, 512<br>CBAM, 512] × 2 |
| Layer5 | --- | --- | BoT Block | BoT Block |
|  | Average Pooling, 2-FC | | | |

We test different models on the public PBC dataset, and all hyperparameters are the same with 30 epochs across the whole training schedule (Table 5). RCMNet has a great improvement compared with the other three models, which indicates that the combination of CBAM and MHSA is effective for image classification for the PBC dataset. 2-FC represents the fully connection with 2 types of output.

Table 5: Comparison among ResNet18, ResNet18-M, ResNet18-C, VGG19, AlexNet, and previous published works and RCMNet under the same setting in Top-1 accuracy and Top-5 accuracy. All images with the resolution of 360 × 363 are trained for 30 epochs. NA is not available.

| Model | Top-1 acc. | Top-5 acc. |
|---|---|---|
| ResNet18 | 99.51 | 100 |
| ResNet18-M | 99.48 | 100 |
| ResNet18-C | 99.58 | 100 |
| RCMNet (Ours) | 99.63(+0.12) | 100 |
| VGG19 | 99.25 | 100 |
| AlexNet | 99.18 | 100 |
| Acevedo *et al.* (*45*) | 96.20 | NA |
| Ucar (*46*) | 97.94 | NA |
| Long *et al.*(*47*) | 99.30 | NA |

## 4.1.2 Model performance comparison with PBC dataset

After evaluating the effectiveness of our proposed model by an ablation experiment, more comparison analyses are carried out. Table 5 shows the test accuracy for all models. The accuracy of AlexNet is the lowest; our model has the highest accuracy. RCMNet improves on top of AlexNet by 0.45% in the regular setting. Here, the accuracy of ResNet18_M is lower than ResNet18 and ResNet18_C, which seemly hints that the CBAM can assist ResNet18 to aggregate features with effective information, achieving better performance after inserting CBAM to the BoT block. Training and testing accuracy, training loss, and confusion matrix of RCMNet classifier after 30 epochs are showed in Figure 8. All models converge after 17 epochs (Figure 8a-c). In Figure 8d, we find that the sub-dataset of IGs shows the most significant confusing score for our model. We speculate that three cell types of promyelocytes, myelocytes, and metamyelocytes belonging to IGs have discernible different phenotypes. Meanwhile, our model achieves higher accuracy with the PCB dataset compared with the previous work.

Figure 8: Training (a) and testing (b) accuracy, training loss (c) and confusion matrix (d) on PBC

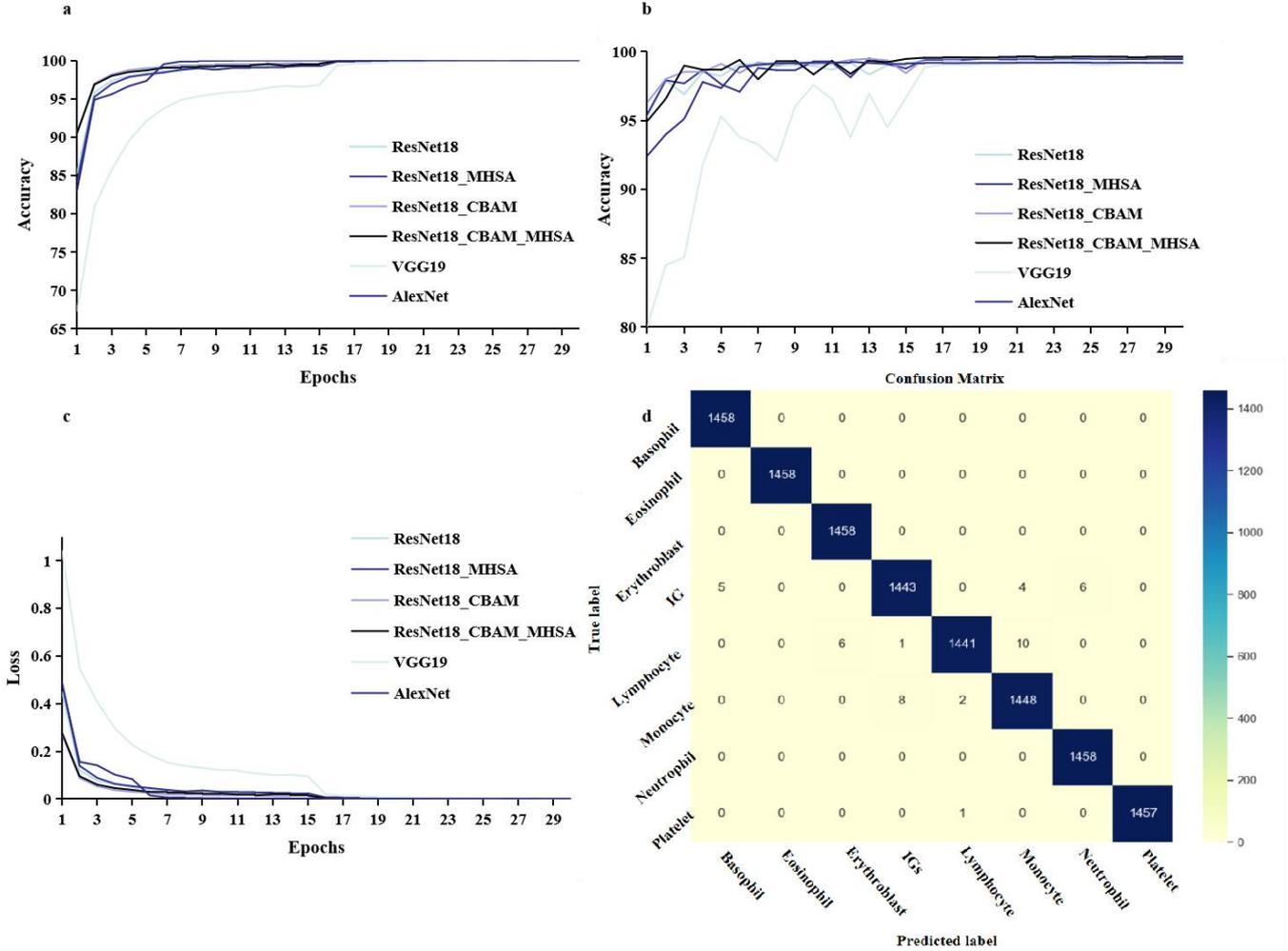

dataset.

## 4.2 Image classification for CAR-T dataset

### 4.2.1 Comparison of the CAR-T dataset

The ablation experiments show promising results on the common dataset with great improvement. Next, we study the performance of RCMNet on the CAR-T dataset. We compare RCMNet model with common models and ResNet18 variants and Top-1 accuracy after 20 epochs are presented in Table 6. Our model shows at least comparable or relatively higher accuracy than most of the models. However, ResNet18 and Resnet18_C outperform than RCMNet in CAR-T dataset. Figure 9 displays the visualization comparison for different models, where RCMNet and ResNet18-M show worse performance, consistent with the calculated results. The focus attention for RCMNet and ResNet18-M is more dispersive. We speculate that because MHSA cannot extract features with enough information for a small dataset causing self-attention incapable of integrating global information comprehensively. Another paper reports a similar conclusion as ours (28). In addition, RCMNet can lose some features in a small dataset, especially for the tiny features during the progress of convolution while they are the key for classification.

Table 6. Comparison of ResNet18, ResNet18-M, ResNet18-C, VGG19, AlexNet, and RCMNet under the same settings in Top-1 accuracy. All images with the resolution of 384 × 384 are trained for 20 epochs.

| Model | Top-1 acc. |
|---|---|
| ResNet18 | 81.24 |
| ResNet18-M | 78.22 |
| ResNet18-C | 82.63 |
| RCMNet | 80.01 |
| VGG19 | 79.11 |
| AlexNet | 78.46 |

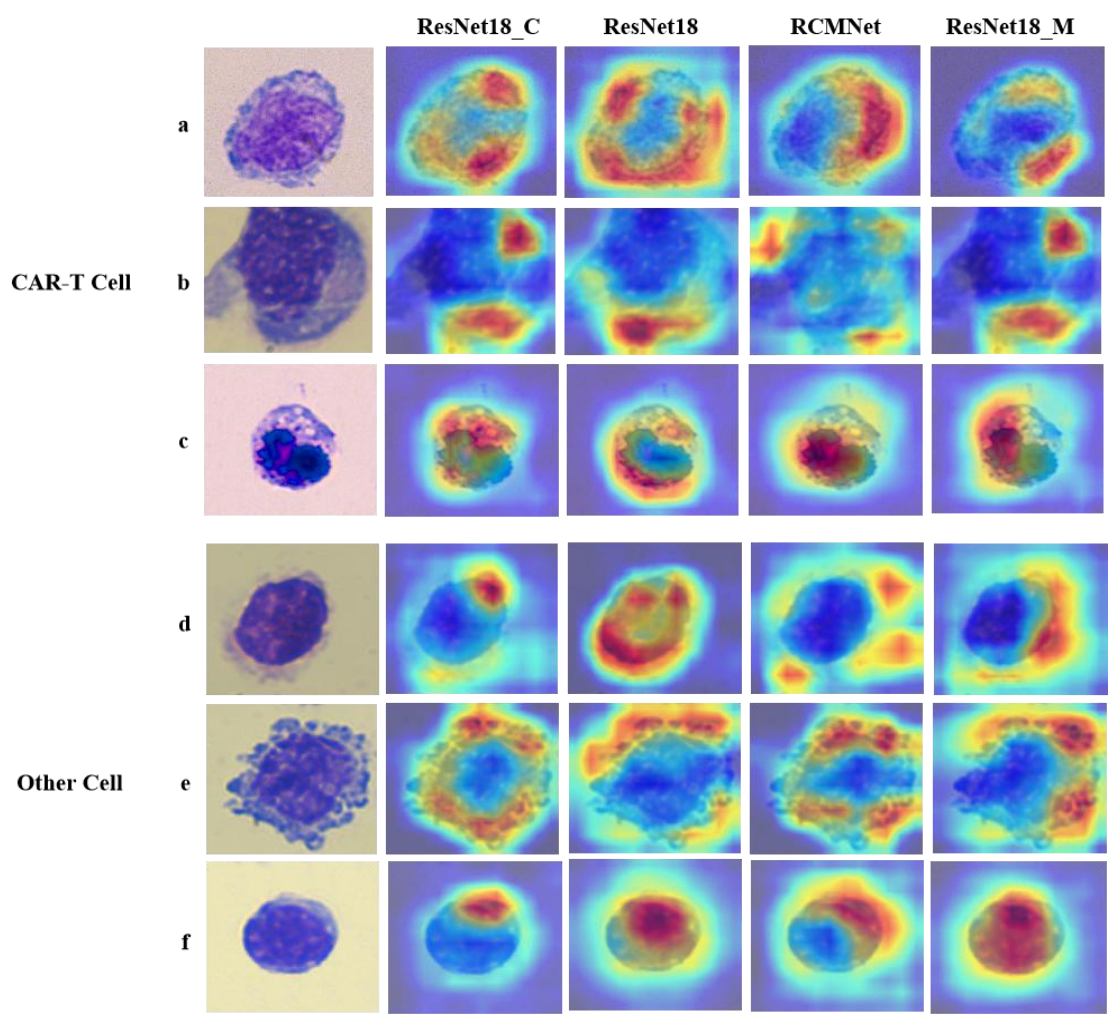

Figure 9 Grad-CAM output from ResNet18, ResNet18-M, ResNet18-C, RCMNet without transfer learning on the CAR-T cell dataset. a-c are the CAR-T cells and d-f are the other blood cells.

### 4.2.2 Enhanced performance of RCMNet via transfer learning

Direct RCMNet model training seemly shows that the accuracy is not ideal due to the limited sample size. The disadvantage and limitations of a small dataset have been discussed extensively since the model cannot get enough training and generalization causing terrible results in predicting unknown cases. Transfer learning can increase the generalization performance between two similar datasets. Table 7 displays the results for each model after transfer learning. In this study, transfer learning is applied to RCMNet and 83.36% accuracy is achieved. Compared with RCMNet without transfer learning, the performance improvement for RCMNet with transfer learning is 3.35%, which is a satisfactory improvement. This result indicates that training RCMNet on CAR-T dataset without transfer learning cannot fully utilizes the power of RCMNet in cell recognition. Figure 10 shows the output of Grad-CAM. RCMNet with transfer learning is easier to focus on the key point with fewer redundant parts. Some failed cases are showed in Figure 11 and the dispersive cases are remaining, which is a challenge for the further study. In addition, the capacity of transfer learning for VGG19, ResNet18, and other models is very limited and the increase is not notable. This result demonstrates the potential of transfer learning in the application of self-attention for medical image processing when large deep learning model is deployed and the small data size is available.

Table 7. Comparison of ResNet18, ResNet18-M, ResNet18-C, VGG19, AlexNet, and RCMNet under the same settings in Top-1 accuracy after transfer learning. All images with the resolution of 360 × 363 are trained for 20 epochs.

| Model | Top-1 acc. |
|---|---|
| ResNet18 | 82.18 |
| ResNet18-M | 82.13 |
| ResNet18-C | 83.01 |
| **RCMNet** | **83.36** |
| VGG19 | 79.69 |
| AlexNet | 79.03 |

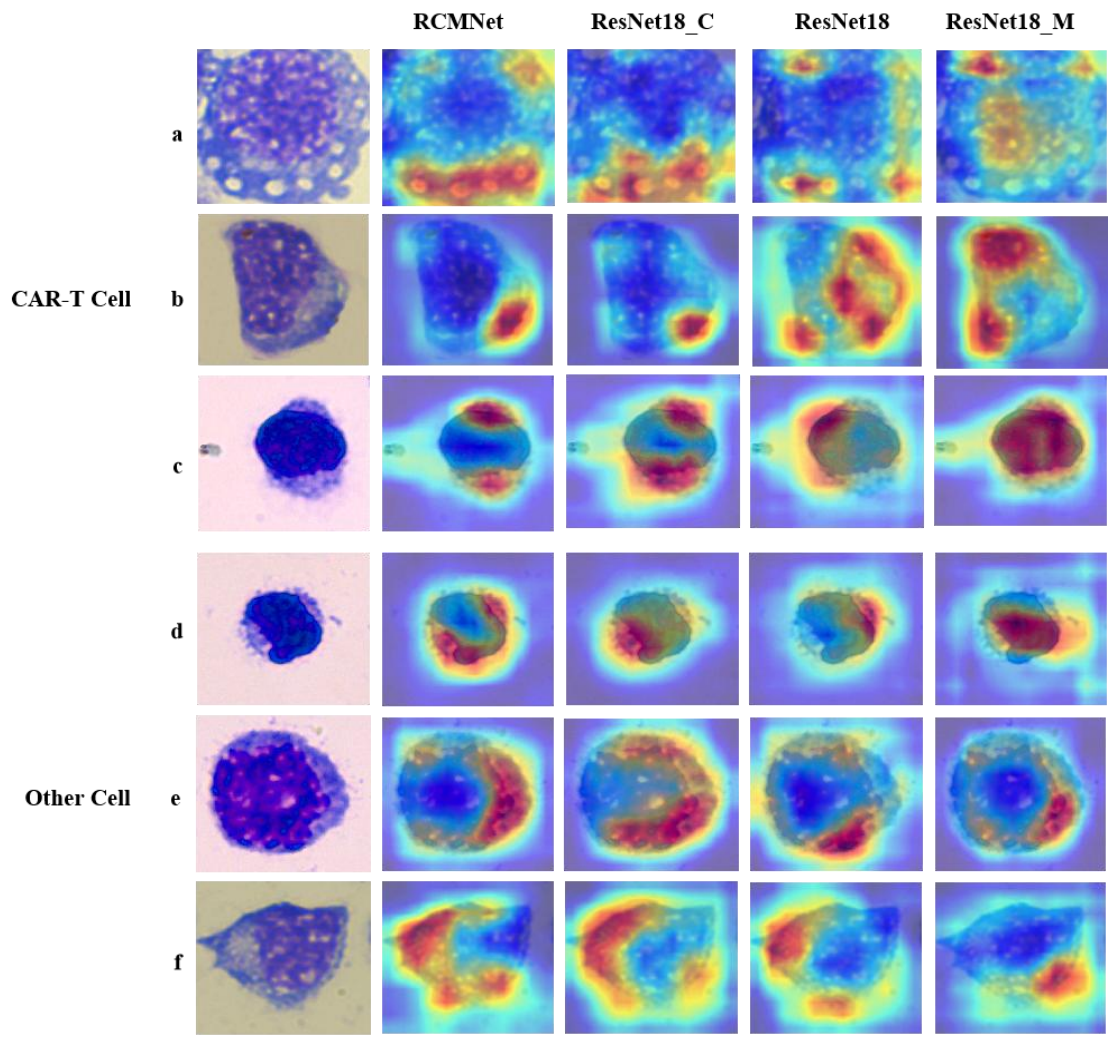

Figure 10: Grad-CAM output from ResNet18, ResNet18-M, ResNet18-C, RCMNet with transfer learning on CAR-T cell dataset. a-c are the CAR-T cells and d-f are the other cells.

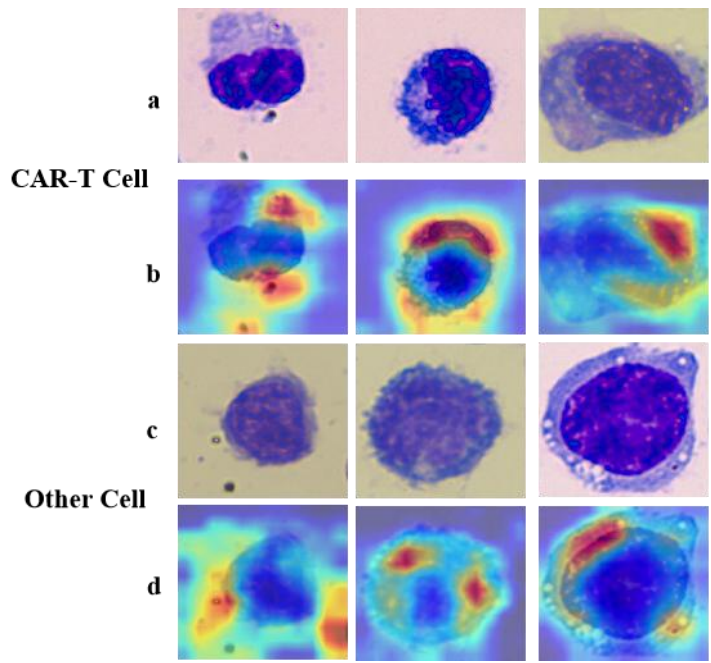

Figure 11: The failed classification cases from RCMNet after transfer learning on the CAR-T cell

dataset. a and c are the original image; b and d are Grad-CAM output.

## 5. Conclusion

Due to the end-to-end inference and efficient feature extraction of deep learning, CAR-T cell recognition is an appropriate and meaningful target due to its clinical value. In this work, we make two unique contributions to the field of CAR-T cell recognition, which still relies on human labor to differentiate the presence of CART cells after CAR-T therapy. (1) We successfully construct a CAR-T dataset containing 500 CART-T cell images, which have been labeled by an experienced blood morphologist. (2) The second one is that we construct a model named RCMNet with two attention mechanisms classifying the CAR-T cell dataset. To the best of our knowledge, we create the first dataset that can be used to develop a classification algorithm for CAR-T cells. Our proposed model RCMNet shows the potential for assisting doctors in identifying CAR-T cell and makes crucial contribution for clinical decision making in the selection of the treatment for acute leukemia. RCMNet is a hybrid model consisting of CNN and self-attention. The biggest benefit is the integration of the local and global information extracted from the images. On the PBC dataset, our model achieves 99.63% top-1 accuracy, outperforming the previous published works. However, the model is sample size-dependent and a large amount of data is preferred. Although the results of our model on the small CAR-T dataset cannot fully exploit its advantage, the accuracy on the single ResNet18-C still achieves 82.63%, and transfer learning seemly a valid method to increase the result up to 83.36% accuracy on the RCMNet. This is the first report on CAR-T cell recognition which is more challenging than regular blood cell classification. The high demand for CAR-T cell therapy urgently requires a platform and method that can facilitate CAR-T cell recognition and clinical decision making. In the future, the CAR-T dataset can be expanded to include more expert labeled images, test the model performance on larger datasets like ImageNet (*48*) and consider replacing the backbone with other architecture, such as inception-v4 (*49*). The complex model architecture and transfer learning on ImageNet can be adapted to improve the model performance in CAR-T cell diagnosis.

## 6. Acknowledgments

This work was supported by grants from National Natural Science Foundation of China (31970752), Science, Technology, Innovation Commission of Shenzhen Municipality (JCYJ20190809180003689, JSGG20200225150707332, WDZC20200820173710001, and JSGG20191129110812708), and Shenzhen Bay Laboratory Open Funding (SZBL2020090501004).